\documentstyle[12pt,fleqn]{article}
\def\mbi#1{\mbox{\boldmath$#1$}}
\def\kp{\mbi{k} \cdot \mbi{p}}

\def\eps{\epsilon}

\def\beeq{\begin{equation}}
\def\eneq{\end{equation}}
\def\beeqa{\begin{eqnarray}}
\def\eneqa{\end{eqnarray}}

\addtocounter{section}{-1}
\begin{document}

\begin{center}

\mbox{}

\mbox{}

{\Large {\bf Impurity Scattering in Carbon Nanotubes
with Superconducting Pair Correlations}}

\mbox{}

\mbox{}

{\large Kikuo Harigaya}

\mbox{}

\mbox{}

{\small {\sl Electrotechnical Laboratory, 
Umezono 1-1-4, Tsukuba 305-8568, Japan}}
\end{center}

\mbox{}

\mbox{}

{\small {\bf 
\noindent
Abstract.}
Effects of the superconducting pair potential on the impurity 
scattering processes in metallic carbon nanotubes are studied 
theoretically.  The backward scattering of electrons vanishes
in the normal state.  In the presence of the superconducting 
pair correlations, the backward scatterings of electron- and
hole-like quasiparticles vanish, too.  The impurity gives rise 
to backward scatterings of holes for incident electrons, and
it also induces backward scatterings of electrons for incident
holes.  Negative and positive currents induced by such the 
scatterings between electrons and holes cancel each other.
Therefore, the nonmagnetic impurity does not hinder 
the supercurrent in the regions where the superconducting 
proximity effects occur, and the carbon nanotube is a good 
conductor for Cooper pairs.  Relations with experiments are 
discussed.

\mbox{}
\mbox{}

\begin{center}
{\large {\bf INTRODUCTION}}
\end{center}

\mbox{}

Recent investigations (1,2) show that the superconducting
proximity effect occurs when the carbon nanotubes 
contact with conventional superconducting metals and wires.
The superconducting energy gap appears in the tunneling 
density of states below the critical temperature $T_{\rm c}$.
On the other hand, the recent theories discuss the nature of
the exceptionally ballistic conduction (3) and the absence
of backward scattering (4) in metallic carbon nanotubes
with impurity potentials at the normal states.

In this paper, we study the effects of the superconducting 
pair potential on the impurity scattering processes in 
metallic carbon nanotubes, using the continuum $\kp$ 
model for the electronic states.  We find the absence of 
backward scatterings of electron- and hole-like 
quasiparticles in the presence of superconducting proximity 
effects, and the nonmagnetic impurity {\sl does not hinder the 
supercurrent} in the regions where the superconducting 
proximity effects occur.  Therefore, the carbon nanotube
is a good conductor for Cooper pairs as well as in the
normal state.  This finding is interesting in view of 
the recent experimental progress of the superconducting 
proximity effects of carbon nanotubes (1,2).

\mbox{}

\mbox{}

\begin{center}
{\large {\bf IMPURITY SCATTERING\\ IN NORMAL NANOTUBES}}
\end{center}

\mbox{}

We will study the metallic carbon nanotubes with the
superconducting pair potential.  The model is as follows:
\beeq
H = H_{\rm tube} + H_{\rm pair},
\eneq
$H_{\rm tube}$ is the electronic states of the carbon
nanotubes, and the model based on the $\kp$ approximation 
(4,5) represents electronic systems on the continuum medium.
The second term $H_{\rm pair}$ is the pair potential term
owing to the proximity effect.

The hamiltonian of a graphite plane by the $\kp$ approximation (4,5) 
in the secondly quantized representation has the following
form:
\beeq
H_{\rm tube} = \sum_{\mbi{k},\sigma} \Psi_{\mbi{k},\sigma}^\dagger
E_{\mbi{k}} \Psi_{\mbi{k},\sigma},
\eneq
where $E_{\mbi{k}}$ is an energy matrix:
\beeq
E_{\mbi{k}} =
\left( \begin{array}{cccc}
0 & \gamma (k_x - i k_y) & 0 & 0 \\
\gamma (k_x + i k_y) & 0 & 0 & 0 \\
0 & 0 & 0 & \gamma (k_x + i k_y) \\
0 & 0 & \gamma (k_x - i k_y) & 0 
\end{array} \right),
\eneq
$\mbi{k} = (k_x, k_y)$, and $\Psi_{\mbi{k},\sigma}$ is an 
annihilation operator with four components:
$\Psi_{\mbi{k},\sigma}^\dagger =
(\psi_{\mbi{k},\sigma}^{(1)\dagger},\\
\psi_{\mbi{k},\sigma}^{(2)\dagger},
\psi_{\mbi{k},\sigma}^{(3)\dagger},
\psi_{\mbi{k},\sigma}^{(4)\dagger})$.
Here, the fist and second elements indicate an electron at 
the A and B sublattice points around the Fermi point $K$ 
of the graphite, respectively.  The third and fourth elements 
are an electron at the A and B sublattices around the Fermi 
point $K'$.  The quantity $\gamma$ is defined as $\gamma 
\equiv (\sqrt{3}/2) a \gamma_0$, where $a$ is the bond 
length of the graphite plane and $\gamma_0$ ($\simeq$ 
2.7 eV) is the resonance integral between neighboring 
carbon atoms.  When the above matrix is diagonalized, we 
obtain the dispersion relation $E_\pm = \pm \gamma 
\sqrt{k_x^2 + \kappa_{\nu \phi}^2 (n)}$, where $k_x$ is parallel
with the axis of the nanotube, $\kappa_{\nu \phi} (n) = (2 \pi / L)
(n + \phi - \nu/3)$, $L$ is the circumference length of the 
nanotube, $n$ ($= 0$, $\pm 1$, $\pm 2$, ...) is the index of bands, 
$\phi$ is the magnetic flux in units of the flux quantum, 
and $\nu$ ($= 0$, 1, or 2) specifies the boundary condition 
in the $y$-direction.  The metallic and semiconducting
nanotubes are characterized by $\nu = 0$ and $\nu = 1$ (or 2),
respectively.  Hereafter, we consider the case
$\phi = 0$ and the metallic nanotubes $\nu = 0$.

The second term in Eq. (1) is the pair potential:
\beeq
H_{\rm pair} = \Delta \sum_{\mbi{k}}
(\psi_{\mbi{k},\uparrow}^{(1)\dagger} 
\psi_{-\mbi{k},\downarrow}^{(1)\dagger}
+\psi_{\mbi{k},\uparrow}^{(2)\dagger} 
\psi_{-\mbi{k},\downarrow}^{(2)\dagger}
+\psi_{\mbi{k},\uparrow}^{(3)\dagger} 
\psi_{-\mbi{k},\downarrow}^{(3)\dagger}
+\psi_{\mbi{k},\uparrow}^{(4)\dagger} 
\psi_{-\mbi{k},\downarrow}^{(4)\dagger}
+ {\rm h.c.} )
\eneq
where $\Delta$ is the strength of the superconducting pair
correlation of an $s$-wave pairing.  We assume that the 
spatial extent of the regions where the proximity effect occurs is
as long as the superconducting coherence length.

Now, we consider the impurity scattering in the normal
metallic nanotubes.  We take into account of the single
impurity potential located at the point $\mbi{r}_0$:
\beeq
H_{\rm imp} = I \sum_{\mbi{k},\mbi{p},\sigma}
{\rm e}^{i(\mbi{k}-\mbi{p}) \cdot \mbi{r}_0}
\Psi_{\mbi{k},\sigma}^\dagger \Psi_{\mbi{p},\sigma},
\eneq
where $I$ is the impurity strength.

The scattering $t$-matrix at the $K$ point is
\beeq
t_K = I [ 1 - I \frac{2}{N_s} \sum_{\mbi{k}} G_K(\mbi{k},\omega)]^{-1},
\eneq
where $G_K$ is a propagator of a $\pi$-electron around the
Fermi point $K$.  The discussion about the $t$-matrix at the $K'$ point is 
qualitatively the same, so we only look at the $t$-matrix 
at the $K$ point.  The sum for $\mbi{k}=(k,0)$, which 
takes account of the band index $n=0$ only, is replaced 
with an integral:
\beeq
\frac{2}{N_s} \sum_{\mbi{k}} G_K (\mbi{k},\omega)
= \rho \int d\eps \frac{1}{\omega^2 - \eps^2}
\left( \begin{array}{cc}
\omega & \eps \\
\eps & \omega
\end{array} \right)
\simeq - \rho \pi i {\rm sgn} \omega
\left( \begin{array}{cc}
1 & 0 \\
0 & 1
\end{array} \right),
\eneq
where $\rho= a / 2 \pi L \gamma_0$ is the density of states at the Fermi energy.
Therefore, we obtain
\beeq
t_K = \frac{I}{1 + I \rho \pi i {\rm sgn} \omega}
\left( \begin{array}{cc}
1 & 0 \\
0 & 1
\end{array} \right).
\eneq
The transformation into the energy-diagonal representation 
where the branches with $E = \pm \gamma |k|$ are diagonal 
has the same form of $t_K$.

The scattering matrix $t_K$ in the energy-diagonal representation
is diagonal, and the off-diagonal matrix elements vanish.  This 
means that only the scattering processes from $k$ to $k$ and 
from $-k$ to $-k$ are effective.  The scatterings from $k$ 
to $-k$ and from $-k$ to $k$ are cancelled.  Such the absence 
of the backward scattering has been discussed recently (4).

\mbox{}

\mbox{}

\begin{center}
{\large {\bf IMPURITY SCATTERING WITH SUPERCONDUCTING
PAIR POTENTIAL}}
\end{center}

\mbox{}

We consider the single impurity scattering
when the superconducting pair potential is present.
In the Nambu representation, the scattering $t$-matrix 
at the $K$ point is
\beeq
\tilde{t}_K = \tilde{I} 
[ 1 - \frac{2}{N_s} \sum_{\mbi{k}} 
\tilde{G}_K (\mbi{k},\omega) \tilde{I}]^{-1},
\eneq
where $\tilde{G}_K$ is the Nambu representation of $G_K$ and
\beeq
\tilde{I} =
I
\left( \begin{array}{cccc}
1 & 0 & 0 & 0 \\
0 & 1 & 0 & 0 \\
0 & 0 & -1 & 0 \\
0 & 0 & 0 & -1
\end{array} \right).
\eneq
The sign of the scattering potential for holes is 
reversed from that for electrons, so the minus sign 
appears at the third and fourth diagonal matrix elements.

The sum over $\mbi{k}$ is performed as in the previous 
section, and we obtain the scattering $t$-matrix (with the same form
in the energy-diagonal representation):
\beeq
\tilde{t}_K =
\frac{I}{1+(I \rho \pi)^2}
\left( \begin{array}{cccc}
1+\alpha \omega & 0 & -\alpha \Delta & 0 \\
0 & 1+\alpha \omega & 0 & -\alpha \Delta \\
-\alpha \Delta & 0 & -1+\alpha \omega & 0 \\
0 & -\alpha \Delta & 0 & -1+\alpha \omega
\end{array} \right)
\eneq
where
$\alpha = I \rho \pi i / \sqrt{\omega^2 - \Delta^2}$.

Hence, we find that the off-diagonal matrix elements 
become zero in the diagonal $2\times2$ submatrix.  This
implies that the backward scatterings of electron-line
and hole-like quasiparticles vanish in the presence of
the proximity effects, too.  Off-diagonal $2\times2$ 
submatrix has the diagonal matrix elements whose 
magnitudes are proportional to $\Delta$.  The finite 
correlation gives rise to backward scatterings of the hole 
of the wavenumber $-k$ when the electron with $k$ is incident.
The back scatterings of the electrons with the wavenumber
$-k$ occur for the incident holes with $k$, too.
Negative and positive currents induced by such the two 
scattering processes cancel each other.  Therefore, 
the nonmagnetic impurity {\sl does not hinder the 
supercurrent} in the regions where the superconducting 
proximity effects occur.  This effect is interesting 
in view of the recent experimental progress of the 
superconducting proximity effects (1,2).

\mbox{}

\mbox{}

\begin{center}
{\large {\bf SUMMARY}}
\end{center}

\mbox{}

We have investigated the effects of the 
superconducting pair potential on the impurity 
scattering processes in metallic carbon nanotubes.
The backward scattering of electrons vanishes 
in the normal state.  In the presence of 
the superconducting pair correlations, the backward 
scatterings of electron- and hole-like quasiparticles 
vanish, too.  The impurity gives rise to backward 
scatterings of holes for incident electrons, and
it also induces backward scatterings of electrons 
for incident holes.  Negative and positive currents 
induced by such the scatterings between electrons 
and holes cancel each other.  Therefore, the carbon
nanotube is a good conductor for the Cooper pairs
coming from the proximity effects.

\mbox{}

\mbox{}

\begin{center}
{\large {\bf REFERENCES}}
\end{center}

\mbox{}

\noindent
1. A. Y. Kasumov et al, Science {\bf 284},
1508 (1999).\\
2. A. F. Morpurgo, J. Kong, C. M. Marcus, and H. Dai,
Science {\bf 286}, 263 (1999).\\
3. C. T. White and T. N. Todorov, Nature {\bf 393},
240 (1998).\\
4. T. Ando and T. Nakanishi, J. Phys. Soc. Jpn. {\bf 67},
1704 (1998).\\
5. H. Ajiki and T. Ando, J. Phys. Soc. Jpn. {\bf 62},
1255 (1993).\\

\end{document}